%% file: sigir2024-resource-selection.tex
\documentclass[acmlarge]{acmart}
\usepackage{tabularx}
\usepackage{subfigure}
\usepackage{multirow}
\usepackage{enumitem}
\usepackage{enumitem}
\usepackage{multirow}
\usepackage{array}
\usepackage{enumitem}
\usepackage{nameref}
\usepackage{adjustbox}
\usepackage{xcolor}
\usepackage{tikz}
\usetikzlibrary{positioning}

\newcolumntype{P}[1]{>{\raggedright\arraybackslash}p{#1}}
\newcolumntype{M}[1]{>{\raggedright\arraybackslash\ttfamily}m{#1}}

\AtBeginDocument{%
  }

\setlist{leftmargin=2mm}

\hypersetup{draft}
\begin{document}

\title[ReSLLM: Large Language Models are Strong Resource Selectors for Federated Search]{ReSLLM: Large Language Models are Strong Resource Selectors for Federated Search}

\author{Shuai Wang}
\affiliation{
	\institution{The University of Queensland}
	\streetaddress{}
	\city{Brisbane}
	\state{QLD}
	\country{Australia}}
\email{shuai.wang@uq.edu.au}

\author{Shengyao Zhuang}
\affiliation{
	\institution{CSIRO}
	\streetaddress{}
	\city{Brisbane}
	\state{QLD}
	\country{Australia}}
\email{shengyao.zhuang@csiro.au}

\author{Bevan Koopman}
\affiliation{
	\institution{CSIRO}
	\streetaddress{}
	\city{Brisbane}
	\state{QLD}
	\country{Australia}}
\email{b.koopman@csiro.au}

\author{Guido Zuccon}
\affiliation{
	\institution{The University of Queensland}
	\streetaddress{}
	\city{Brisbane}
	\state{QLD}
	\country{Australia}}
\email{g.zuccon@uq.edu.au}


\begin{abstract}

Federated search, which involves integrating results from multiple independent search engines, will become increasingly pivotal in the context of Retrieval-Augmented Generation pipelines empowering LLM-based applications such as chatbots. These systems often distribute queries among various search engines, ranging from specialized (e.g., PubMed) to general (e.g., Google), based on the nature of user utterances. 

A critical aspect of federated search is resource selection - the selection of appropriate resources prior to issuing the query to ensure high-quality and rapid responses, and contain costs associated with calling the external search engines. However, current SOTA resource selection methodologies primarily rely on feature-based learning approaches. These methods often involve the labour intensive and expensive creation of training labels for each resource. In contrast, LLMs have exhibited strong effectiveness as zero-shot methods across NLP and IR tasks. We hypothesise that in the context of federated search LLMs can assess the relevance of resources without the need for extensive predefined labels or features. 

In this paper, we propose ReSLLM. Our ReSLLM method exploits LLMs to drive the selection of resources in federated search in a zero-shot setting. In addition, we devise an unsupervised fine tuning protocol, the Synthetic Label Augmentation Tuning (SLAT), where the relevance of previously logged queries and snippets from resources is predicted using an off-the-shelf LLM and then in turn used to fine-tune ReSLLM with respect to resource selection. 
Our empirical evaluation and analysis details the factors influencing the effectiveness of LLMs in this context. The results showcase the merits of ReSLLM for resource selection: not only competitive effectiveness in the zero-shot setting, but also obtaining large when fine-tuned using SLAT-protocol.



\end{abstract}


\keywords{Large Language Models, Ranking, Prompt}

\maketitle

\input{sections/introduction.tex}

\input{sections/related-works.tex}

\input{sections/methods.tex}
\input{sections/experimental_setup.tex}

\input{sections/results.tex}

\input{sections/conclusion.tex}


\bibliographystyle{ACM-Reference-Format}
\bibliography{sigir2024-resource-selection}

\end{document}

%% file: sections/introduction.tex
\section{Introduction}


\textit{Federated search} is a technique that integrates search results from multiple search engines or databases, offering a comprehensive and unified view of the information landscape to the user~\cite{shokouhi2011federated}. This approach is particularly valuable in scenarios where information is dispersed across various sources, each with its unique structure and content specialization, e.g. Amazon focuses on products, while Arxiv focuses on scientific publications. By leveraging federated search, users can execute a single query across these diverse sources, receiving aggregated results that represent a broader spectrum of relevant data. 
A critical component of federated search is \textit{resource selection}. This process involves determining which search engines (resources) to query based on the user's search intent (expressed by the query). 
Effective resource selection is pivotal in federated search as it directly influences the relevance and quality of the information retrieved~\cite{shokouhi2011federated}. Furthermore, effective resource selection also reduces the time and costs associated with API querying~\cite{nguyen2012federated,nguyen2016resource}: only a subset of resources need to be queried, rather than all of them. These selected resources are deemed most likely to contain the information pertinent to the user's query.

 Resource selection and federated search are critical for modern chatbot agents based on Retrieval-Augmented Generation (RAG), such as langchain\footnote{\url{https://www.langchain.com/}}, llamaindex~\cite{Liu_LlamaIndex_2022}, DSPy~\cite{khattab2022demonstrate,khattab2023dspy}. These LLM-based agents integrate a retrieval module within the response generation process, enhancing the reliability and accuracy of the outputs, and reducing the model hallucination problem, which commonly appears in pre-trained large language models. This retrieval component often is required to access multiple resources (indexes, search engine APIs). Appropriate resource selection is then critical for the reasons listed above, and because only a handful of results could be handled by the pipeline if the aggregation is performed by a LLM, due to the high latency typically associated with this component. 
 
Effective resource selection methods predominantly rely on human-annotated datasets (i.e. with relevance judgements)~\cite{dai2017learning,wu2019ltrrs,ergashev2023learning}; obtaining these annotations is both labor-intensive (thus costly) and time-consuming, especially in enterprise and domain-specific settings (e.g. in health). These methods often utilize learning algorithms dependent on manually annotated labels to correlate user queries with resource relevance. 

This paper breaks with previous resource selection methods in that it devises state-of-the-art methods that do not require the creation of human labels.
 For this, we explore the application of Large Language Models (LLMs)
to select resources for federated search systems. The motivation for doing this is intuitive: LLMs have recently exhibited remarkable capabilities across various tasks, including question answering~\cite{shao2023prompting}, summarization~\cite{zhang2023benchmarking,yang2023exploring}, and passage ranking~\cite{zhuang-etal-2023-open, sachan-etal-2022-improving, zhuang2023beyond,wang2024zero, wang2024zero, wang2023generating}, even without human-labelled training data. Thus we posit we can effectively adapt them to the resource selection task. 
To this aim, we devise ReSLLM, a prompting method for LLMs that relies solely on the search engine name/description to represent a resource and decide upon its selection.
ReSLLM can be executed as a \textit{zero-shot} approach, i.e. without the need for any training. In addition, we devise a fine-tuning protocol for ReSLLM: the Synthetic Label Augmentation Tuning (SLAT). This protocol consists in acquiring logged queries previously submitted to all resources, along with the search results returned by each resource. Then an off-the-shelf LLM is used to create synthetic relevance judgements (labels) for each search result. Labels are then aggregated at a resource level and used to estimate the relevance of each resource. This information is used to \textit{fine-tune} ReSLLM: labels do not involve human intervention, and supervision is provided by the separate LLM.

%
%


%

To investigate the effectiveness of ReSLLM, also in conjunction with the SLAT protocol, we instruct experiments using the publicly available TREC FedWeb 13 and 14 collections. In our analysis we explore the factors that influence ReSLLM effectiveness, including LLM model size and architecture, resource representation, query types, and the use of SLAT fine-tuning. Our findings reveal that the zero-shot ReSLLM prompting, when configured with specific backbone LLMs, outperforms the current state-of-the-art unsupervised and embedding-based baselines in terms of effectiveness. Notably, when augmented with the SLAT-tuning protocol, ReSLLM demonstrates a performance comparable to the SOTA supervised baselines, which typically depend on human-generated labels for model tuning. This is particularly significant as the SLAT protocol operates effectively without the necessity for such training labels, indicating a promising direction for resource selection in federated search environments. Our finding also show that while the selection of size and architecture of the backbone model attribute significantly for the zero-shot ReSLLM, the importance diminishes when the model is fine-tuned with SLAT.


%% file: sections/related-works.tex
\section{Related Work}

\subsection{Resource Selection}
\citet{shokouhi2011federated} and \citet{garba2023federated} offer a foundational understanding of federated search, highlighting the challenges of selecting from diverse information sources. 
Current resource selection methods in federated search are broadly classified into unsupervised and supervised approaches.

\textbf{Unsupervised Methods.}
Unsupervised methods refer to methods for which no human labels are needed to inform the model on how to select resources: this method is further categorized into lexicon-based and sample-based methods. Lexicon-based approaches~\cite{callan2002distributed,xu1999cluster}, view resources as large lexicons, employing document word statistics for resource representation.
On the other hand, sample-based algorithms, including ReDDE \citep{si2003relevant}, CRCS \citep{shokouhi2007central}, and SUSHI \citep{thomas2009sushi}, utilize a Centralized Sample Index (CSI) from sampled documents. The latest in this category, KBCS \citep{han2018knowledge}, enhances query-resource similarity assessment using contextually enriched entity sets. Despite their effectiveness, as noted by \citet{ergashev2023learning}, these methods may not fully address the structural intricacies of textual inputs.
\citet{garba2020embedding} introduced a paradigm shift towards semantic analysis in federated search with their model based on word embeddings, demonstrating the growing emphasis on semantic congruence in resource selection.

\textbf{Supervised Methods.}
Supervised approaches in federated search leverage machine learning techniques for resource selection, ranging from query and resource classification to learning-to-rank methods. Models like SVMrank~\cite{dai2017learning} and LambdaMART-based LTRRS~\citep{wu2019ltrrs} optimize rankings using diverse features, including term statistics and CSI-based metrics. However, these methods often require extensive feature engineering, as highlighted by their limitations in capturing a wide array of query and resource features.

\citet{ergashev2023learning} proposed a graph neural network learning-to-rank method. The approach uses a  graph to encapsulate query-resource and resource-resource relationships, and it leverages pre-trained language models for enhanced semantic understanding. The method however is heavily reliant on training data, with relevance assessments used for training sourced from the same dataset and it is thus unsuitable for the task we consider in this paper.

\textbf{Other Methods.} 
Resource selection shares similarities with shard selection, with previous studies on shard-based selective search offering a different perspective on resource selection~\cite{kulkarni2015selective,mendoza2016reducing}.

\subsection{Language Models in Information Retrieval}
We employ Large Language Models (LLMs) in our work to judge and rank documents. A wealth of studies in NLP and IR has consistently showcased the impressive zero-shot capabilities of LLMs. In particular, studies have illustrated the robust zero-shot capabilities of LLMs in ranking documents. The methodologies for harnessing LLMs in zero-shot ranking tasks encompass Pointwise~\cite{zhuang-etal-2023-open, sachan-etal-2022-improving, zhuang2023beyond,wang2024zero}, Listwise~\cite{sun-etal-2023-chatgpt, ma2023zero, pradeep2023rankvicuna, pradeep2023rankzephyr, tamber2023scaling}, Pairwise~\cite{qin2023large}, and Setwise~\cite{zhuang2023setwise} approaches. In our work, we adapt the Pointwise method to rank resources providing user query using the logits probability of the next generated token to be `yes' or `no'.

On the other hand, research by \citeauthor{guglielmo2023perspectives}\cite{guglielmo2023perspectives} suggests that LLM-generated relevant labels can closely align with labels annotated by TREC annotators.  Similarly, findings by \citeauthor{thomas2023large}\cite{thomas2023large} suggest that utilizing GPT-4 for generating relevant judgments can outperform human annotators in evaluating search systems. We adapt this method in our work to judge relevance of documents.

%% file: sections/methods.tex
\section{Methodology}
\label{method}

\begin{figure}[t!]
	\centering
	\includegraphics[width=0.9\textwidth]{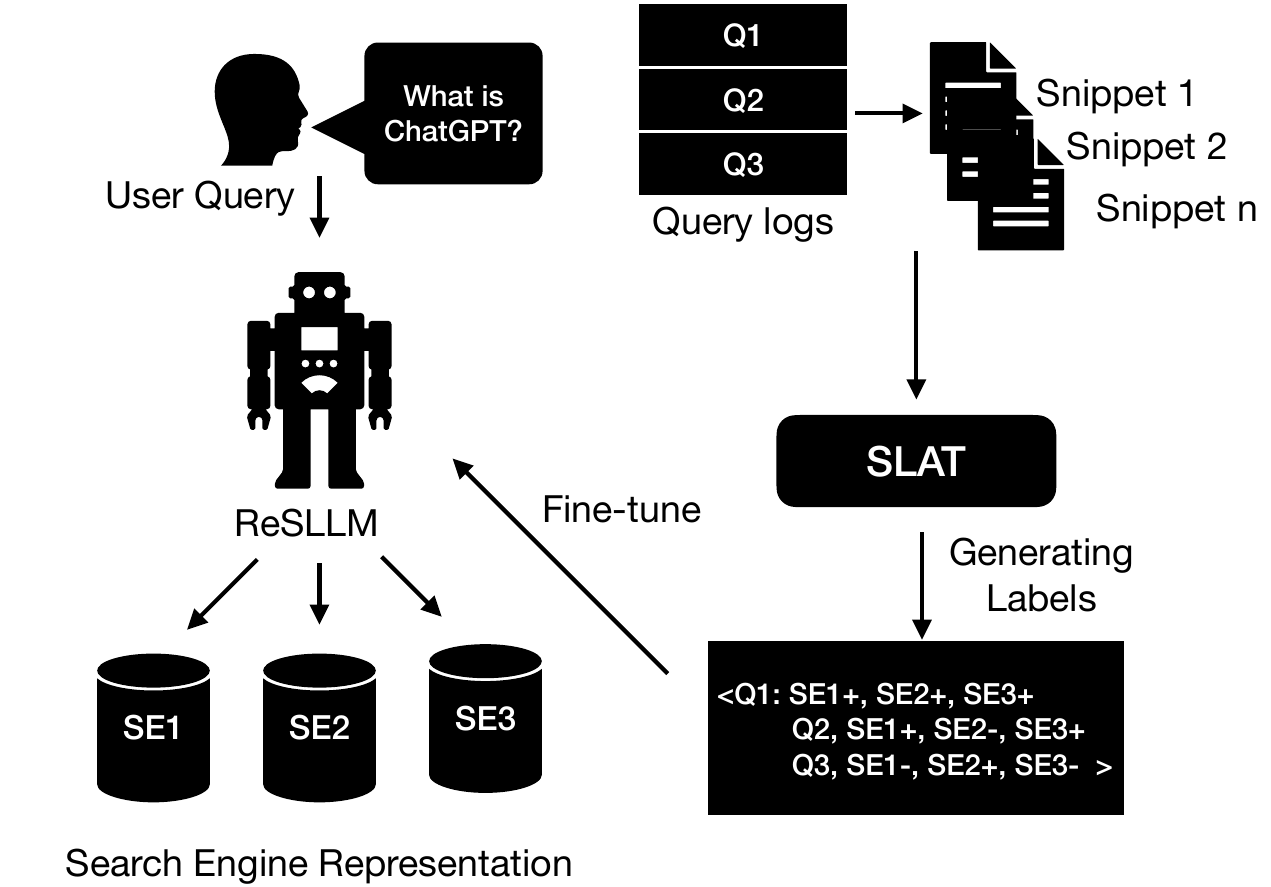}

	\caption{The use of ResLLM for resource selection, including the SLAT protocol for fine-tuning. }
	\label{fig:architecture}
\end{figure}

ReSLLM leverages LLMs as a text-based resource selector to identify the resources for which a query should be routed to in a federated search setting. Figure~\ref{fig:architecture} provides an overview of the use of ReSLLM for resource selection, including the choices available in terms of resource representation and fine-tuning (the SLAT protocol).

\subsection{Problem Formulation}
The resource selection problem for federated search is defined as follows. Given a set $R$ containing $n$ resources ($R = \{r_1, \ldots, r_n\}$) and a user query $q$, an effective resource selector $\mathcal{S}$ would select only a suitable subset of resources $S(q)$ that best answer $q$, i.e. resources that contain relevant documents for the query. In our context, resources are search engines, although this could be further extended to other type of information access systems like databases, knowledge bases, etc.. 

However, in practice search engine practitioners implementing a federated search engine might have additional constrains on the maximum number of search engines the resource selector $\mathcal{S}$ should provide, e.g. only the top-$k$ resources should be returned by the selector. We note this restriction is particularly important in RAG pipelines where the results of the federated search are then handed over to a LLM for synthesis. In many implementations of these pipelines we have examined, in fact, the merging of results from the resources is performed by the LLM. 
In this case, in fact, a large value of $k$ would correspond to high latency (as each merge incurs a costly LLM inference), and thus a small $k$ value is often preferred. For example, the RAG pipeline may consider using the downstream LLM to merge the top $m=5$ search results from the top $k=3$ search engines. Other times the LLM would keep merging search results and create a synthesis on-the-fly until certain condition is met, e.g., the generated answer is complete or does not change further. To address these settings, we reformulate the resource selection problem in that the subset of resource $S(q)$ returned by the resource selector $\mathcal{S}$ for query $q$ needs to be ordered in decreasing likelihood of relevance (i.e. ranked). The evaluation measures we use to evaluate methods for resource selection will provide an understanding of their effectiveness with respect to both selecting a subset of $k$ resources, and ranking such a subset so that the most relevant resources are at the top (see Section~\ref{sec:eval_measures}).

We consider another restriction to the above setup. In most RAG pipelines, as is often the case in other applications of federated search, the resources involved in the federation are \textit{uncooperative}, i.e., they do not provide any information about their content to the resource selector, aside from a description of the search service and the research results obtained in answer to queries~\cite{shokouhi2011federated,nguyen2012federated,urak2018source}. Thus, statistical properties of the underlying collection indexed by a resource are not available. It is possible however to maintain or acquire a log $\mathcal{L}$ of previously submitted queries (\textit{logged queries}) and the associated search snippets returned by each resource.

\subsection{The ReSLLM Prompting Method}
In ReSLLM, we employ an instruction fine-tuned Large Language Model to address the resource selection task, without the need for additional task-specific fine-tuning involving expensive and often impractical collection of labelled data. This approach involves directly inputting the user query $q$ and the resource representation (e.g., resource name) in the prompt, which the LLM uses to answer a yes/no question. This decision determines the appropriateness of the chosen search engine for the specified query. 

The ReSLLM prompt follows the format [\textit{System Instruction}, \textit{Resource Representation}, \textit{User Input}, \textit{Task Instruction}], and is shown in  Table~\ref{table:prompts}. The System Instruction spells out the context of the task, defining the federated search setting, the user query, the objective of the resource selection task. The Resource Representation provides the information that is used to represent a specific resource; these can be the name of the search engine (\textit{\{name\}}), the URL of the search engine (\textit{\{url\}}), the description of the search engine (\textit{\{description\}}). The User Input is used to pass the user's query (\textit{\{query\}}) and (\textit{\{similar\_sampled\_snippets\}}) from the search engine for which the LLM needs to select resources. 
Finally, the Task Instruction includes text that describes the expected output from the LLM.


\begin{table}[t!]
	\centering
	\small
	\caption{The ReSLLM prompt. In curly brackets are resource and input variables, e.g., the name of the search engine and the user's query. In the experiments we study the effect of including (or not) the description/snippets. The parts of the prompt that are added (removed) when (not) considering description/snippets are marked in underlined text. }
		\resizebox{0.95\columnwidth}{!}{
		\begin{tabular}{p{2pt}|M{420pt}}
			\toprule
			& \textnormal{Prompt} \\
			\hline
			\rotatebox[origin=c]{90}{System Instruction} & Federated search retrieves information from a variety of sources via a search application built on top of one or more search engines. A user makes a single query request. The federated search then selects only the search engines that the query should be sent to from a list of search engines, and aggregates the result for presentation of high quality result to the user. The task is called resource selection. \\ \midrule
			
			\rotatebox[origin=c]{90}{{\centering Resource Representation}} & The following is a search engine with its name, url \underline{and description}.\linebreak
			Name: $\{name\}$\linebreak
			URL: $\{url\}$\linebreak
			\underline{Description: $\{description\}$} \\ \midrule
			
			\rotatebox[origin=c]{90}{{\centering User Input}} &
			The following is a real user query: \linebreak
			Query: $\{query\}$ \linebreak
			\underline{The following are some snippets from this search engine that are similar }\linebreak
			\underline{to the user query:} \linebreak
			\underline{Snippets: $\{similar\_sampled\_snippets\}$} \linebreak \\ \midrule
			
			\rotatebox[origin=c]{90}{{\centering Task Instruction}} &
			Now, please reply only yes or not to indicate if the query should be sent to the search engine.\linebreak
			Response:\\
			\bottomrule
		\end{tabular}
	}
	\label{table:prompts}
\end{table}

Once the ReSLLM prompt is executed on an LLM, we record the logits associated with the tokens "yes" and "no". Then, for each resource $r_i$, we use the logits to obtain the probabilities $P(\text{yes}|q, r_i)$ and $P(\text{no}|q, r_i)$, via soft-max over all tokens' logits. ReSLLM then uses the following equation to score a resource:

\begin{equation}
	\label{eq:zero-shot-decision}
	score(q, r_i) = P(\text{yes}|q, r_i) - P(\text{no}|q, r_i)
\end{equation}

Resources are then ranked by descending $score(q, r_i)$. In the empirical experiments in this paper we report the effectiveness of the rankings obtained according to equation~\ref{eq:zero-shot-decision}; note however that in practice one might restrict the selected resources only to those with $score(q, r_i) \ge 0$ (if they are less than $k$), as resources with a negative scores are more likely to be non-relevant than relevant.

\subsection{The SLAT Protocol}

We are interested in applications of federated search, e.g., in the context of a RAG pipeline, in which acquiring relevance labels from human assessors is not possible or is inconvenient -- for example due to data privacy, cost or time-to-deployment. 

While we assume that in such contexts document-level relevance labels are not available, at times the firm implementing the resource selection module might  have access to a query log $\mathcal{L}$, containing previous queries submitted to the federation of search engines, or example queries akin to those expected from users, along with the top $m$ search result snippets obtained from each resource.
The availability of such a query log motivates us to devise a protocol in which a LLM is used to derive synthetic relevance labels for each pair of query and search snippet. These labels are aggregated for resource on a query base, so as to obtain a relevance score for each resource. These are then used to fine-tune the resource selector $\mathcal{S}$. We name this protocol for fine-tuning as Synthetic Label Augmentation Tuning (SLAT). Next we describe the four components of SLAT: the automatic relevance assessment, the aggregation of relevance to the resource level, the creation of the synthetic training labels, and the actual fine-tuning of ReSLLM.

\begin{table}[t!]
	\centering
	\small
	\caption{Prompt used for instructing the LLM to generate relevance assessments; prompt adapted from \cite{thomas2023large}.	}
	\begin{tabular}{p{2pt}|M{420pt}}
		\toprule
		& \textnormal{Prompt} \\
		\hline
		\rotatebox[origin=c]{90}{System Prompt} &  Given a query and a web snippet, you must provide a score on an integer scale of 0 to 4 with the following meanings:\linebreak 
		4 = navigational, this page represents a home page of an entity directly named by the query; the user may be searching for this specific page or site.\linebreak 
		3 = top relevance, this page or site is dedicated to the topic; authoritative and comprehensive, it is worthy of being a top result in a web search engine.\linebreak  
		2 = highly relevant, the content of this page provides substantial information on the topic.\linebreak 
		2 = highly relevant, the content of this page provides substantial information on the topic.\linebreak 
		1 = minimal relevance, the content of this page provides some information on the topic, which may be minimal; the relevant information must be on that page, not just promising-looking anchor text pointing to a possibly useful page.
		\linebreak 0 = not relevant, the content of this page does not provide useful information on the topic, but may provide useful information on other topics, including other interpretations of the same query. \\ \midrule
		
		\rotatebox[origin=c]{90}{{\centering User Prompt}} & Assume that you are writing a report on the subject of the topic. mark the web snippet according to the previous scale description.
		\linebreak Query:\linebreak A person has typed \{query\} into a search engine.
		\linebreak Result:\linebreak Consider the following web snippet:
		\linebreak —BEGIN WEB Snippet CONTENT—
		\linebreak \{snippet\}\linebreak
		—END WEB Snippet CONTENT—
		\linebreak Instructions:\linebreak 
		Split this problem into steps:
		\linebreak Consider the underlying intent of the search.\linebreak 
		Measure how well the content matches a likely intent of the query (M)\linebreak 
		Measure how trustworthy the web page is (T).\linebreak 
		Consider the aspects above and the relative importance of each, and decide on a final score (O).\linebreak Produce a JSON of scores without providing any reasoning. Example:\{"M": 2, "T": 1,"O": 1\}\linebreak 
		Results:\\
		\bottomrule
	\end{tabular}
	\label{table:prompt_label}

\end{table}


\subsubsection{Automatic Relevance Assessment}
Our SLAT protocol for fine-tuning relies on a LLM to create synthetic (pseudo) relevance assessments for a query-snippet pair from the query log $\mathcal{L}$. Recent work from \citet{thomas2023large} has shown LLMs are effective for relevance assessments, and the MS Bing search engine does reportedly utilise synthetic relevance labels created using GPT-4.0 for training. Within SLAT, we rely on an adaptation of the prompt used by \citet{thomas2023large}, and on an open-source backbone LLM (detailed in Section~\ref{sec:setup}), rather than the close-source (and expensive) GPT-4.0 they used. The original prompt used a three-level relevance scale; we instead employ a five-level relevance scale to match the scale used in the collections we shall consider in our experiments (from the TREC FedWeb Track): Navigational, Key, Highly Relevant, Relevant, and Non-Relevant. Table~\ref{table:prompt_label} provides our prompt and definitions for each level of relevance. In the original study,~\citet{thomas2023large} investigate many variations of the prompt; we choose the most effective prompt they presented that adhere to our inputs, i.e., the prompt that activates only the aspects module.

\subsubsection{Aggregation of Relevance to the Resource Level}
After obtaining synthetic assessments for individual query-snippet pairs for a resource, the relevance of the resource for that  given query is determined by calculating the graded precision of the top 10 results~\cite{demeester2013overview}. This in all effects results in an estimation of the true graded precision based on the relevance level predictions provided by the LLM. For computing the graded precision of a resource given the query-snippet assessments, we use the same weights used by the evaluation framework of the TREC FedWeb 2013 Track : \( w_{\text{Non}} = 0 \) (non-relevant), \( w_{\text{Rel}} = 0.25 \) (relevant), \( w_{\text{HRel}} = 0.5 \) (highly relevant), and \( w_{\text{Key}} = w_{\text{Nav}} = 1 \) (key and navigational). Note that this choice is made to align our method to the collection in which it is tested in this paper; but the choice of value can be adapted to the context in which SLAT and ReSLLM are used, without further repercussions on the methods.
The graded relevance values of the query-snippet pairs are then transformed into discrete relevance levels, through multiplication by 100 and rounding to the nearest integer, again following the practice in TREC FedWeb 2013 Track. This aggregation results in a relevance score between 0 to 100, with 100 indicating the highest relevance score for a resource, and 0 indicating the lowest score.

\subsubsection{Creation of Synthetic Training Labels}
The next step involves transforming the aggregated relevance status of each resource for a query in the log $\mathcal{L}$ into actionable training data for fine-tuning the LLM backbone of ReSLLM. For this, we adopt an instruction-based approach, categorizing resources into three distinct intervals based on their relevance: 

\begin{itemize}
	\item \textit{Highly Relevant Resources:} These are resources where the aggregated relevance scores are indicative of a high degree of relevance of the resource to the logged query. For fine-tuning purposes, these resources are labeled with two 'yes' responses, denoting their strong alignment with the query.
	
	\item \textit{Marginally Relevant Resources:} Resources falling into this category exhibit some level of relevance but are not closely aligned to the logged query. For these, we use a mixed labeling approach, consisting of one 'yes' and one 'no'.
	
	\item \textit{Not Relevant Resources:} This category encompasses resources that are deemed not relevant or of little value in relation to the logged query. Correspondingly, these resources are labeled with two 'no' responses in our training data, indicating their lack of relevancy.
\end{itemize}

\subsubsection{ReSLLM Fine-tuning} 
Finally, we use our training data generated through the execution of the previous three steps to fine-tune the backbone LLM or ReSLLM. For this, we use the prompt in Table~\ref{table:prompts}, where we populate the query value with the logged query. Then we used a contrastive loss to function to fine-tune our downstream LLMs with instruction-based training samples.


%% file: sections/experimental_setup.tex
\section{Experimental Setup}
\label{sec:setup}

\subsection{Considered LLMs}

We consider seven LLMs in our experiment, comprising a mix of encoder-decoder and decoder-only models (to study effect of LLM architecture on ReSLLM) and model size (to study the effect of LLM size on ReSLLM):

\begin{itemize}
	\item \textit{Flan-T5 Models~\cite{flan-t5}:} Enhanced versions of Google's T5 models, Flan-T5 models have undergone instruction-based fine-tuning for various generation tasks. We experiment with the flan-t5-large (783 million), flan-t5-xl (2.8 billion), and flan-t5-xxl (11.3 billion), with the flan-t5-xl being the default model we use across most experiments.


	\item \textit{LlaMa2 Models~\cite{touvron2023llama2}:} An extension of the LlaMa family, LlaMa2 models are pre-trained on 2 trillion tokens, representing a 40\% increase from LlaMa~\cite{touvron2023llama,touvron2023llama2}. Notably, LlaMa2 includes a specialized “Chat” variant, fine-tuned for chat-based interactions and capable of following user instructions. We experiment with LlaMa2-7b-chat (7 billion) and LlaMa2-13b-chat (13 billion) models.

\end{itemize}

In addition to the models above, we also selected two of the latest, high-performing, open-source LLMs from the huggingface open\_llm\_leaderboard\footnote{\url{https://huggingface.co/spaces/HuggingFaceH4/open_llm_leaderboard}}. These two models were selected based solely on their average effectiveness in reducing perplexity across the various NLP tasks in the leaderboard at the time of our experiments. We specifically focused on models with up to 13 billion parameters, the upper limit dictated by our computational resources. Other model-specific characteristics, such as architecture or size, were not considered in this selection process. The models selected under these criteria are:
	\begin{itemize}

	\item \textit{Mixtral\_7Bx2\_Moe}\footnote{\url{https://huggingface.co/cloudyu/Mixtral_7Bx2_MoE}}: A 12.9 billion parameter mixture of experts LLM, combining the capabilities of two 7B sub-models.

	\item \textit{SOLAR-10.7B-Instruct-v1.0}\footnote{\url{https://huggingface.co/upstage/SOLAR-10.7B-Instruct-v1.0}}: Employs a depth up-scaling technology, as described by~\citet{kim2023solar} from the Mistral-7B model to obtain a 10.7 billion parameter model.


\end{itemize}

\subsection{Datasets}
We use the two TREC Federated Web track collections distributed as part of the \textit{FedWeb Greatest Hit}~\cite{demeester2015fedweb}:  \textit{FedWeb-2013 (FedWeb-13)} and \textit{FedWeb-2014 (FedWeb-14)}.


\textit{FedWeb-13} incorporates data from 157 distinct in-production search engines (resources), providing a comprehensive basis for the resource selection tasks. It contains 50 test queries, each with detailed descriptions and narratives. 
Included along with the queries, there are also the top 10 search snippets retrieved by each of the 157 search engines for each of the test queries. Relevance assessments are provided for all these query-snippet pairs.
Additionally, FedWeb-13 includes snippets extracted from the top ten documents returned for 2,000 distinct synthetic queries (single=term queries extracted from snippets), offering a representative sample of each search engine's content~\cite{demeester2013overview}.

\textit{FedWeb-14} incorporates data from 149 search engines (less than in FedWeb-13 due to the 8 being discontinued). It also provides test queries with descriptions and narratives, along with top-10 retrieved snippets and associated relevance assessments. Complementing the collection are an additional 4,000 single-term synthetic queries, with associated top-10 snippets for each search engine~\cite{demeester2014overview}.


FedWeb Greatest Hit includes an additional 406 unique queries with the corresponding top-10 snippets from each search engine; these queries and snippets are not used in FedWeb-13/2014 and no relevance assessments for them are available.
In our experiments, we collate these 406 queries and associated snippets, along with the (unlabbled) single-term queries and associated snippets from FedWeb-13/2014 to form a \textit{query log}. We then use this query log  to investigate the effectiveness of conducting SLAT-based fine-tuning for our ReSLLM. We also use this query log to create a competitive unsupervised embedding baseline, described in Section~\ref{sec:baselines}.

In our experiments we consider two types of test queries: ad-hoc queries and conversational queries; conversational queries are considered because of our specific-interest in resource selection in the context of modern RAG pipeline, which are increasingly used within conversational agent-based language tools, like chatbots. For ad-hoc queries we use the queries provided by the FedWeb collections. For conversational queries, we use the descriptions associated to the queries in FedWeb.

\subsection{Evaluation measures}
\label{sec:eval_measures}


Our evaluation primarily utilizes normalized Discounted Cumulative Gain (nDCG) at cut-offs 10, 20 and 100, and normalized Graded Precision (nP) at cut-offs 1 and 5 , following common practice in federated search and TREC FedWeb collections.
nP evaluates the relevance of the top-k resources~\cite{demeester2013overview}. It calculates the Graded Precision at $k$ (nP@k) based on graded relevance weights. nP@k is then normalized by the highest Graded Precision achievable by any resource for that topic at the same cut-off. 

\subsection{Baselines}
\label{sec:baselines}

Our primary baselines are derived from the submitted runs of TREC FedWeb 2013 and 2014. The baselines were selected based on specific criteria to ensure a comprehensive and fair evaluation:

\begin{enumerate}
	\item \textbf{Supervised Baseline:} The supervised baseline relies on relevance assessments provided by human annotators, which are in turn used to train a learning model. The baseline is chosen based on the best nDCG@20 performance (primary metric) from the TREC leaderboard. From TREC FedWeb 2014, the run \textit{ECNUCS-ecomsvz}~\cite{jin2014simple} is selected as supervised baseline. This run is notable for not only relying on human-annotated labels but also for incorporating external APIs (Google search) and being a fused run combining various approaches submitted. No run is selected from TREC FedWeb 2014, as no run is supervised.

	\item \textbf{Unsupervised Baseline:} The unsupervised baseline is the highest performing method from the TREC leaderboard that neither uses human labels for training nor relies on additional external API services. From TREC FedWeb 2013, \textit{UPD-UPDFW13mu} is chosen as unsupervised baseline; this baseline implements the TWF-IRF weighting scheme for resource selection~\cite{di2014university}. From TREC FedWeb 2014, \textit{Dragon-drexelRS7} is selected as unsupervised baseline; this run uses the SUSHI method, which uses additional data from the query log (i.e. the documents pointed by the snippets retrieved from the logged queries) to represent resources and derive selection scores for each search engine~\cite{zhao2014drexel}.
\end{enumerate}

We note that both \textit{ECNUCS-ecomsvz} and \textit{Dragon-drexelRS7} use additional data our ReSLLM method does not have access to (commercial market share, external Google API, sampled documents etc); In particular, \textit{ECNUCS-ecomsvz}, as the supervised baseline in 2014, also used annotated labels from test set in TREC FedWeb to select resource. We decided to include them nevertheless as the strongest baselines from the corresponding TREC tracks.

Additionally, we devise and implement an Embedding baseline, leveraging recent advancements in pre-trained embedding models, particularly dense retrievers~\cite{zhao2022dense, wang2023balanced}. This baseline uses the BAAI/bge-large-en-v1.5 model~\cite{bge_embedding} to encode queries and resources. The BAAI/bge-large-en-v1.5 model~\cite{bge_embedding} model is one of the top encoding models in the MTEB leaderboard~\cite{muennighoff2022mteb} at the time of experimentation, showcasing strong zero-shot performance across domains. To represent resources (search engines), we consider the snippets from the query log associated to each resource. For each resource, we encode the snippets from the query log using the BAAI/bge-large-en-v1.5 model. At inference time, we use the model to encode the user queries; the cosine similarity between the encoding of resources' snippets and the query encoding is then computed, and resources are scored with the average of the scores obtained by the top-3 snippets for the query on that resource.


\subsection{SLAT Settings}

\subsubsection{LLM used in SLAT for synthetic relevance assessments generation}
SLAT uses an LLM to generate relevance assessments for logged queries and snippets using the prompt in Table~\ref{table:prompt_label}. Queries, snippets and synthetic assessments are then used within the SLAT protocol to obtain training labels for the resources, which are in turn used for instruction-based fine-tuning of ReSLLM. We undertook some preliminary experiments to understand which LLM we could have used to obtain the synthetic relevance assessments. We could not rely on the LLM used by \citet{thomas2023large}, GPT-4.0, as we would have incurred prohibitive costs and experiments would have taken an excessive long time due to the numerous API calls. We then considered open-source LLMs. Assessments produced by flan-t5-xl, flan-t5-xxl  and llama2-7b-chat proved being of low quality. We instead succeeded in producing meaningful relevance assessments with the SOLAR-10.7B-Instruct-v1.0, as thus we adopted this model as LLM for synthetic relevance generation. As a point of reference, we generated relevance assessments for the snippets already assessed by TREC assessors in FedWeb13 and FedWeb14 (two assessors), and binarised grouping together Nav, Key and HRel into a single Relevant category. We then computed inter-rater agreement using Cohen's kappa statistic, and found an agreement of $\kappa=0.35$ between the LLM and each of the assessor; the agreement between the two assessors is $\kappa=0.6$.

\subsubsection{Training queries for SLAT \& generation of conversational queries}
When using SLAT, we need access to training queries. When fine-tuning ReSLLM for ad-hoc queries, we use the  queries supplied in the query log. However, we cannot reliably use the logged queries when fine-tuning ReSLLM for conversational queries, as most of the logged queries are single-term queries, while the conversational queries are long descriptions.

We then use the SOLAR-10.7B-Instruct-v1.0 LLM to generate conversational queries from the logged queries and the associated logged search snippets. For this, we first use the LLM to provide a relevance judgement for each snippet for a logged query, using the same process described above. We then sampled for each logged query 3 snippets that the LLM has assessed as Navigational relevant.

From here, we devise a prompt that instructs the LLM to generate a conversational query given the ad-hoc query, the 3 sampled snippets and an example query pair <\textit{ad-hoc}, \textit{conversational}> we manually  constructed (to avoid using test data for the example). The prompt in shown in Table~\ref{table:qgen_prompt}.

\begin{table}[t!]
	\centering
	\small
	\caption{Prompt used for generating a conversational query from an ad-hoc query and 3 navigationally relevant snippets. }
	\resizebox{0.9\columnwidth}{!}{
	\begin{tabular}{p{2pt}|M{420pt}}
		\toprule
		& \textnormal{Prompt} \\
		\hline
		\rotatebox[origin=c]{90}{System Instruction} & Given a user query and relevant snippets about the query, a description of the query describes the user information need with respect to the relevant snippets. \\ \midrule

		\rotatebox[origin=c]{90}{{\centering Example}} &
		An example query is: protege pizza tutorial\linebreak
		\linebreak
		then the relevant snippets are:\linebreak
		\linebreak
		Snippet 1:\linebreak
		Snippet Title: In annual tradition, advertisers cowed by NFL trademark bullying.\linebreak
		Snippet Info:\linebreak
		... to sell a variety of products$\backslash$u2014televisions, pizzas, soda$\backslash$u2014in conjunction with the game... $\backslash$"Super Sale XLVI,$\backslash$" using a football for a logo. Pizza Hut is offering a $\backslash$"Big Deal…\linebreak
		\linebreak
		The description of the user query is:\linebreak
		You are looking for a tutorial or guide related to making or preparing pizza, possibly with a focus on specific techniques or styles. This could include step-by-step instructions, tips, or video demonstrations.\\ \midrule

		\rotatebox[origin=c]{90}{{\centering Task Instruction}} &
		Now, given the following user query:\linebreak
		\{query\}\linebreak
		the relevant snippets are:\linebreak
		\{snippets\}\linebreak
		Please write a description for the user query above.\\
		\bottomrule
	\end{tabular}
}

	\label{table:qgen_prompt}
\end{table}


\subsubsection{Fine-tuning Settings for SLAT}
\label{sec:fine-tune_setting}
To create synthetic training labels, we used three aggregated score intervals to describe the relevance level of a resource to a query. For scores $\ge50$ -- \textit{Highly Relevant Resources}; scores between 25 to 50 --  \textit{Marginally Relevant Resources}, and for scores $<25$ -- \textit{not Relevant Resources}
Finally, we fine-tuned three LLMs using SLAT: Flan-large, Flan-xl, and LlaMa2-7b-chat. These models were selected based on the computational resources available to us.
For the Flan-large and Flan-xl models, fine-tuning was performed using the PyTorch Lightning library~\cite{Falcon_PyTorch_Lightning_2019}. We conducted the fine-tuning over two epochs with a contrastive loss function. The batch sizes were set to 20 for Flan-large and 5 for Flan-xl, this size was chosen to maximize the utilization of our GPU memory. Except for the batch size, all other parameters were set to their default values in the library.
Regarding the LlaMa2-7b-chat model, it was fine-tuned following the standard instructions for the LlaMa2 chat model. This fine-tuning was carried out for one epoch, with a batch size of 2 and gradient accumulation steps set to 5. For other parameters, we adhered to the fine-tuning guidelines same  to those in the Stanford-Alpaca project~\cite{alpaca}.

%% file: sections/results.tex
\begin{table*}[t]
	\centering
	\resizebox{\textwidth}{!}{
	\begin{tabular}{l|ll|lllll|lllll}
		\toprule
		&&&\multicolumn{5}{c|}{Ad-hoc} &\multicolumn{5}{c}{Conversational} \\ \cmidrule(lr){4-13}
		&Method&LLM&ndcg10&ndcg20&ndcg100& nP1&nP5&ndcg10&ndcg20&ndcg100& nP1&nP5 \\ \midrule
		\multirow{7}{*}{\rotatebox{90}{FedWeb-13}}
		&\multicolumn{2}{l|}{Unsupervised Baseline}  & - & 0.2990& - &0.1600&0.2100&- & - & - & - &\\   
		& \multicolumn{2}{l|}{Embedding Baseline} & 0.2067 & 0.2508 & 0.5088 & 0.1611 & 0.1880 & 0.2789 & 0.3398 & 0.5608 & 0.2259 & 0.2481 \\ \cmidrule(l){2-13}
		& ReSLLM & flan-xl  & 0.1511* & 0.2054* & 0.4123* & 0.1239 & 0.1374* & 0.1789* & 0.2143* & 0.4384* & 0.1939 & 0.1638* \\
		& ReSLLM & llama2-7b-chat  & 0.0683* & 0.0815* & 0.2626* & 0.0273* & 0.0644* & 0.1156* & 0.1467* & 0.3599* & 0.0780* & 0.1045* \\
		& ReSLLM & solar-11b  &  \textbf{0.2636*} &  0.3029 &  \textbf{0.5351} &  \textbf{0.2713*} &  0.2486& \textbf{0.3573*} & \textbf{0.3944} & \textbf{0.5982} & \textbf{0.3809*} & \textbf{0.3398} \\
		& ReSLLM & mixtral-13b  & 0.2605 &  \textbf{0.3059} & 0.5224 & 0.1897 &  \textbf{0.2492} & 0.2802 & 0.3246 & 0.5562 & 0.2392 & 0.2726 \\
		& ReSLLM & llama2-13b-chat  & 0.1015* & 0.1339* & 0.3036* & 0.1045 & 0.1069* & 0.1640* & 0.2140* & 0.4129* & 0.1733 & 0.1564* \\ \midrule
		
		\multirow{8}{*}{\rotatebox{90}{FedWeb-14}}&\multicolumn{2}{l|}{Unsupervised Baseline}  &0.3590&0.4220& - &0.2930&0.3140& - & - & - & - & -\\
		
		&\multicolumn{2}{l|}{Supervised Baseline}  &\textbf{0.6240}& \textbf{0.7120}& - & \textbf{0.5350}&\textbf{0.6040}& - & - & - & - & - \\ 
		& \multicolumn{2}{l|}{Embedding Baseline}& 0.1928 & 0.2418 & 0.5041 & 0.1940 & 0.1740 & 0.2680 & 0.3364 & 0.5568 & 0.2271 & 0.2489 \\ \cmidrule(l){2-13}
		& ReSLLM & flan-xl  & 0.1484* & 0.1845* & 0.3799* & 0.1097 & 0.1377 & 0.1801* & 0.2092* & 0.4087* & 0.1717 & 0.1866 \\
		& ReSLLM & llama2-7b-chat  & 0.0529* & 0.0717* & 0.2500* & 0.0232* & 0.0441* & 0.0606* & 0.0976* & 0.3132* & 0.0355* & 0.0594* \\
		& ReSLLM & solar-11b  & 0.4093* & 0.4227* & 0.6115* & 0.3931*& 0.3926* & 0.4351*& 0.4509*& 0.6362*& 0.3803* & 0.4363*\\
		& ReSLLM & mixtral-13b  & 0.2870 & 0.3431 & 0.5614 & 0.2896 & 0.2572 & 0.2927 & 0.3558 & 0.5680 & 0.2980 & 0.2653 \\
		& ReSLLM & llama2-13b-chat  & 0.0977* & 0.1184* & 0.3370* & 0.1047 & 0.0856* & 0.1599* & 0.1945* & 0.4008* & 0.1987 & 0.1401* \\
		
		\bottomrule
	\end{tabular}
}
	\caption{Effectiveness of zero-shot ReSLLM method in comparison with baseline approaches. $*$ indicates statistical significant differences against the Embedding Baseline (two-tail paired t-test, $p<0.05$). Effectiveness for TREC baselines for conversational queries is not attainable, as the track did not consider these queries.}
	\label{table:main_table}
\end{table*}

\section{Results and Analysis}


\subsection{Zero-shot ReSLLM}
\label{sec:res_zero_shot}

For these experiments, we use the name and url to represent resources in the prompt of ReSLLM; we investigate the impact of other resource representations in Section~\ref{sec:impact_representation}. For the Flan-T5 model family, we only report the results obtained with flan-xl in this section. The Flan-T5 model family allows us to study three LLM sizes, so we focus on the effect of model size for the Flan-T5 model family in a separate analysis (Section~\ref{sec:model_size}).



Table~\ref{table:main_table} presents a comparative analysis of our zero-shot ReSLLM method against the baseline approaches. 

When comparing the effectiveness across different LLM backbones, the results highlight a significant dependency of ReSLLM's zero-shot effectiveness on the chosen backbone. In particular, the Solar-11b model generally outperforms other LLMs across most metrics, with the exception of ndcg@20 and nP@5 for FedWeb-13 when ad-hoc queries are used: in this case, Mixtral-13b exhibits marginally higher effectiveness.

When compared to the Embedding baseline, the use of Solar-11b and Mixtral-13b demonstrates significant improvements across most evaluation metrics in both collections. However, it's worth noting that the baseline Embedding model exceeds the effectiveness of other LLMs (flan-xl, LlaMa2-7b-chat, and Llama-2-13b chat).

Against unsupervised baselines, the zero-shot ReSLLM with Solar-11b consistently achieves a higher effectiveness. However, any of the other backbone exhibits lower effectiveness across all metrics in comparison with supervised baselines. 
The supervised baseline (\textit{ECNUCS-ecomsvz}) requires annotated labels from human judgment and external commercial API support. Unlike this baseline, results for ReSLLM are solely based on prompting the LLM in a zero-shot manner.

Furthermore, the results show that ReSLLM performs better when queries are conversational, compared to when queries are of the ad-hoc type; this is expected as conversational queries contain more information about the user needs that the ReSLLM can interpret to rank resources. 

In summary, ReSLLM typically surpasses both the Unsupervised and Embedding baselines in terms of effectiveness for selected LLM backbones, albeit not outperforming current supervised state-of-the-art models trained directly on human labels.



\subsection{ReSLLM with SLAT}
In this section we investigate the effectiveness of the SLAT protocol for fine-tuning ReSLLM without resorting to human relevance assessments. 
As in Section~\ref{sec:res_zero_shot}, also here we only consider the representation of resources based only on name and url for the prompt of ReSLLM. In addition, we only consider fine-tuning the ReSLLM model based on flan-xl. Fine-tuning other larger LLMs (flan-xxl and the decoder-only LLMs, apart from llama2-7b-chat) require computational resources that exceed those we have available. We report results obtained when fine tuning flan-large and llama2-7b-chat when studying the effect of LLM size and architecture (Section~\ref{sec:model_size}).

\begin{table*}
	\centering
		\resizebox{\textwidth}{!}{
	\begin{tabular}{l|ll|lllll|lllll}
		\toprule
		&&&\multicolumn{5}{c|}{Ad-hoc} &\multicolumn{5}{c}{Conversational} \\ \cmidrule(lr){4-13}
		&Method&LLM&ndcg10&ndcg20&ndcg100& nP1&nP5&ndcg10&ndcg20&ndcg100& nP1&nP5 \\ \midrule
		\multirow{5}{*}{\rotatebox{90}{FedWeb-13}}
		&\multicolumn{2}{l|}{Unsupervised Baseline}  & - & 0.2990& - &0.1600&0.2100& - & - & - & - & -\\  
		& \multicolumn{2}{l|}{Embedding Baseline} & 0.2067 & 0.2508 & 0.5088 & 0.1611 & 0.1880 & 0.2789 & 0.3398 & 0.5608 & 0.2259 & 0.2481 \\ \cmidrule(l){2-13}
		& ReSLLM &flan-xl  & 0.1511 & 0.2054 & 0.4123 & 0.1239 & 0.1374 & 0.1789 & 0.2143 & 0.4384 & 0.1939 & 0.1638 \\
		& SLAT+ReSLLM & flan-xl  & \textbf{0.7557*} & \textbf{0.7897*} & \textbf{0.8574*}  & \textbf{0.6929* } & \textbf{0.7752*}  & \textbf{0.7556*}  & \textbf{0.7988*} & \textbf{0.8622*} & \textbf{0.6950*} & \textbf{0.7857*} \\\midrule 
		
		\multirow{6}{*}{\rotatebox{90}{FedWeb-14}}&\multicolumn{2}{l|}{Unsupervised Baseline}  &0.3590&0.4220& - &0.2930&0.3140& - & - & - & - & -\\
		&\multicolumn{2}{l|}{Supervised Baseline}  &0.6240& \textbf{0.7120}& - & \textbf{0.5350}&0.6040& - & - & - & - & - \\ 
		& \multicolumn{2}{l|}{Embedding Baseline} & 0.1928 & 0.2418 & 0.5041 & 0.1940 & 0.1740 & 0.2680 & 0.3364 & 0.5568 & 0.2271 & 0.2489 \\  \cmidrule(l){2-13}
		& ReSLLM &flan-xl   & 0.1484 & 0.1845 & 0.3799 & 0.1097 & 0.1377 & 0.1801 & 0.2092 & 0.4087 & 0.1717 & 0.1866 \\
		& SLAT+ReSLLM & flan-xl  & \textbf{0.6582*} &0.7088* & \textbf{0.7855*} & 0.3060* & \textbf{0.6854*} & \textbf{0.6557*} & 0.7052*& \textbf{0.7833*} &0.3060*& \textbf{0.7008*} \\
		
		\bottomrule
	\end{tabular}
}
	\caption{Effectiveness of SLAT+ReSLLM, in comparison to the zero-shot ReSLLM and the baselines. $*$ indicates statistical significant differences against the Embedding Baseline (two-tail paired t-test, $p<0.05$). Effectiveness for TREC baselines for conversational queries is not attainable, as the track did not consider these queries.}
	\label{table:tuned_table}
\end{table*}

\begin{figure*}[t!]
	\centering
	\includegraphics[width=\textwidth]{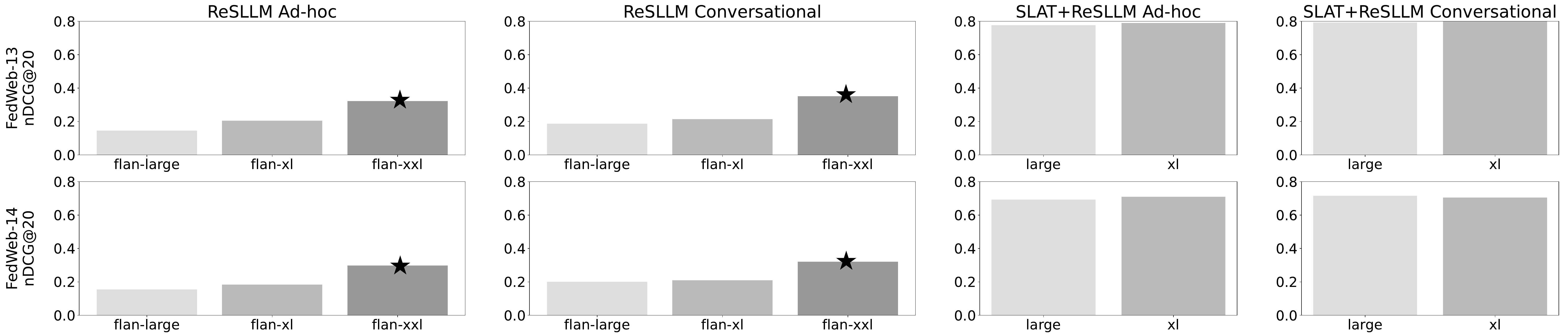}

	\caption{Comparison of effectiveness with respect to the size of model in terms of ndcg@20. Flan-based models are used across all setting. $*$ indicates statistical significant differences against the flan-large model (two-tail paired t-test, $p<0.05$). }
	\label{fig:size}
\end{figure*}

We can use the results in Table~\ref{table:tuned_table} to compare SLAT+ReSLLM vs. the zero-shot version of ReSLLM and the baselines. The results highlight the benefits in fine-tuning ReSLLM with the SLAT protocol: this setup achieves by far the highest effectiveness across all baseline methods, except for some metrics when compared to the supervised baseline in FedWeb-14. 

The significant gap in nP1 between SLAT+ReSLLM and the supervised baseline in FedWeb-14 is surprising, especially in light of the higher nP5 SLAT+ReSLLM obtains.
We attribute this to our SLAT fine-tuning protocol, and in particular to the choice made when setting the threshold set for pseudo-label creation (Section~\ref{sec:fine-tune_setting}). This may lead to a flattening of scores in the top of the rankings, affecting the differentiation necessary to obtain higher nP@1 scores. However, in practical scenarios, such as within RAG pipelines where results are aggregated and re-ranked from diverse sources, this limitation may be less concerning. The re-ranking process in these applications is likely to provide a more varied input to the LLM, potentially offsetting the observed experimental limitations.

\begin{figure*}[t!]
	\centering
	\includegraphics[width=\textwidth]{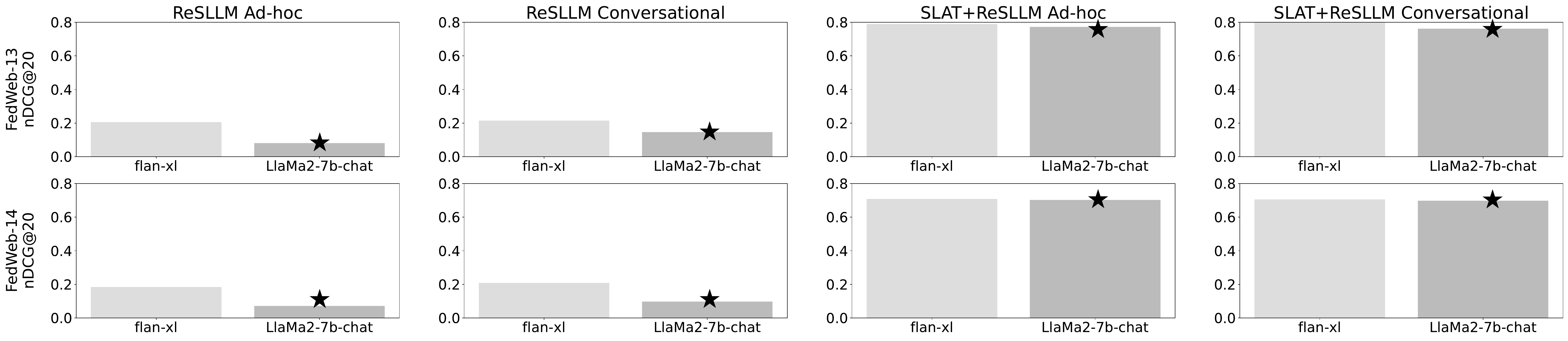}

		\caption{Comparison of effectiveness with respect to the architecture of the model. $*$ indicates statistical significant differences against the flan-xl model (two-tail paired t-test, $p<0.05$). }
	\label{fig:archi}
\end{figure*}

In our analysis, we found no significant improvement when applying SLAT+ReSLLM to conversational queries as compared to ad-hoc queries. This observation may also be attributed to our SLAT fine-tuning approach. Specifically, our methodology involved transforming ad-hoc queries from the query log into conversational queries, which could inadvertently alter the original intent of the queries. As a result, the labels generated by the LLM responsible for relevance assessments might not accurately reflect the modified conversational context.
We recognize this as a limitation of our current study. The potential mismatch in query intent and the corresponding labels necessitates a more nuanced approach to fine-tuning for conversational queries. In future work, we plan to explore this aspect in greater depth, potentially involving the re-judgment of labels by the LLM to better align with the conversational query context. This will help in understanding the distinct dynamics of conversational search and improving the efficacy of models like SLAT+ReSLLM in such scenarios.

\begin{figure*}[t!]
	\centering
	\includegraphics[width=\textwidth]{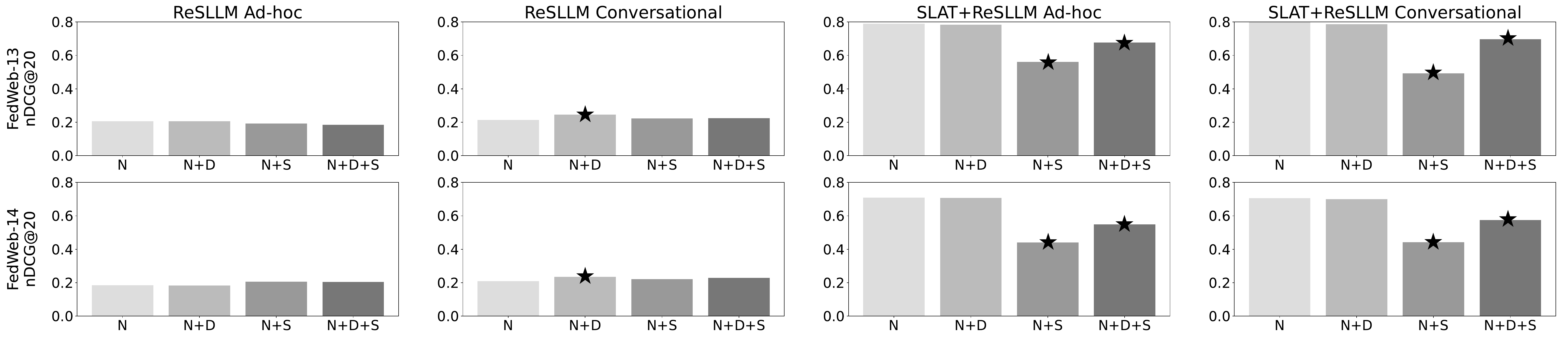}

	\caption{Comparison of effectiveness with respect to the resource representation in terms of ndcg@20. $N$ indicates Name; $D$ indicates Description; $S$ indicates Similar snippet. flan-xl is used across all setting. $*$ indicates statistical significant differences against $N$ for each setting. (two-tail paired t-test, $p<0.05$). }
	\label{fig:source_representation}
\end{figure*}

In summary, we find that the SLAT protocol brings substantial improvements to ReSLLM, unsurprisingly. SLAT+ReSLLM not only outperforms all unsupervised and embedding baselines but also achieves effectiveness on par with the supervised baseline, which relies on human-annotated data, commercial APIs, and method fusion. This is in contrast to SLAT, that does not depend on human annotations; instead it uses LLM-generated labels for informing the fine-tuning process. This highlights SLAT's potential as an effective tuning protocol for enhancing ReSLLM, especially in scenarios where human-annotated data is limited or unavailable.

\subsection{Impact of LLM Size and Architecture}
\label{sec:model_size}
Next we consider how characteristics of the LLM like size ( number of parameters) and architecture (encoder-decoder vs. decoder-only) impact the zero-shot ReSLLM and the SLAT fine-tuned version.

\vspace{2pt}
\textbf{Impact of Model Size.} For this we fix the architecture, comparing the effectiveness of the three models in the Flan-T5 family. 
Figure~\ref{fig:size} illustrates the comparative effectiveness of these models for both the zero-shot ReSLLM and SLAT+ReSLLM. 
We observe a clear trend for the zero-shot ReSLLM: an increase in the size of the LLM significantly enhances the effectiveness of ReSLLM. For instance, nDCG@20 increases by an average of 32\% from flan-large (0.8b parameters) to flan-xl (2.8b parameters), and by a further 60\% when moving to flan-xxl (11.8b parameters). 

However, this disparity in performance diminishes once the models are fine-tuned (SLAT+ReSLLM). Notably, we observe a negligible difference in resource selection effectiveness between flan-large and flan-xl when fine-tuned using SLAT (average increase of 1\%). This finding suggests that while the size of an LLM plays a significant role in zero-shot resource selection, the advantages of larger model sizes decrease after fine-tuning. Note for the SLAT+ReSLLM we could not test flan-xxl because fine-tuning this model exceeded the capabilities of our computational resources.

\vspace{2pt}
\textbf{Impact of Model Architecture.} 
Figure~\ref{fig:archi} compares the performance of different model architectures, focusing on 
flan-xl (an encoder-decoder model) and LlaMa2-7b-chat (a decoder-only model). In the zero-shot ReSLLM setup, flan-xl demonstrates significantly higher effectiveness than LlaMa2-7b-chat, particularly in terms of ndcg@20. However, this performance disparity narrows markedly after SLAT fine-tuning, suggesting that fine-tuning effectively mitigates the differences attributable to inherent architectural and pre-training characteristics for downstream tasks.

This observation implies that architectural features are pivotal in a zero-shot context, but their impact is considerably reduced following specialized fine-tuning, like SLAT for resource selection. Thus, when selecting a backbone LLM for zero-shot resource selectors, careful consideration is warranted, ideally with preliminary tests to assess the model's performance. Conversely, post fine-tuning with protocols like SLAT, the choice of LLM may become less critical due to the alignment and optimization achieved through fine-tuning.


\subsection{Impact of Resource Representation}
\label{sec:impact_representation}
To systematically analyze the impact on resource selection effectiveness of using different resource representation modalities in ReSLLM's prompt (Table~\ref{table:prompts}), we experiment with three distinct types of resource representations: name including the resource url (N), description (D), and similar snippets (S). names and urls of resources are provided in the FedWeb collections. To obtain descriptions, we create a resource description ourselves by searching Wikipedia for information about the resource and extracting the first two sentences from the associated page. In cases no Wikipedia page for the resource is available, descriptions are sourced from the resource's official website. When resources do not exist anymore (e.g. the Disney Family resource), we search the web to investigate the information that resource contained, and write a description that mimics the format of those extracted from Wikipedia. 
These descriptions provide an overview of the resources. The complete set of descriptions, alone with code, is available in the GitHub website associated to this paper\footnote{\url{https://anonymous.4open.science/r/SLAT-RsLLM-F232/} }. To derive the similar snippets representation, for each resource, we use the top 3 snippets from the logged snippets obtained using in the Embedding baseline.


The impact of resource representation on the effectiveness of ReSLLM is shown in  Figure~\ref{fig:source_representation}. 
In the zero-shot setting, enriching source representation modestly enhances resource selection effectiveness. Specifically, the effectiveness improves by approximately 6\% when combining name and description (N+D), and  3.5\% for name and similar snippets (N+S); The integration of all three elements (N+S+D) provides only marginal improvements (+1.1\%).

In contrast, when employing SLAT tuning, the effectiveness varies more substantially across different source representations. Interestingly, the combination of name and description (N+D) shows a negligible difference in effectiveness (less than 1\%). However, the inclusion of similar snippets (N+S) leads to a significant decrease in effectiveness ($\approx-35$\%). Combining all elements (N+S+D) mitigates this decline to some extent, thought it still is less effective (-16\%).


These findings suggest that while richer representations can enhance the effectiveness of ReSLLM for resource selection in a zero-shot setting, this advantage diminishes after the model is SLAT fine-tuned. Particularly, the inclusion of similar snippets in the prompt significantly impacts effectiveness post-tuning. This highlights the importance of carefully considering the combination of resource representations in different operational settings of ReSLLM.


%% file: sections/conclusion.tex
\section{Conclusion}
This study investigated the effectiveness of utilizing Large Language models for resource selection in federated search. 
Our study has three primary contributions. First, we introduce ReSLLM, a zero-shot LLM-based method for resource selection that operates without the need for human-labeled data. Second, we develop the Synthetic Label Augmentation Tuning (SLAT) protocol, a fine-tuning approach that leverages LLM-generated synthetic labels, providing an effective means of tuning ReSLLM without human intervention. Third, through extensive experiments addressing our research questions, we demonstrate that both ReSLLM and SLAT+ReSLLM can effectively select resources in a federated search environment, outperforming several baseline methods and matching the efficacy of supervised models in certain scenarios.

Our contributions have the potential to significantly benefit RAG pipelines, especially for conversational agents. The ability to accurately select resources has profound implications for improving the quality of downstream generation tasks. While our study did not directly explore this specific impact, due to lack of evaluation resources that permit this, it lays the groundwork for future research in this area: one important direction being the creation of federated search datasets that enable the evaluation of this technology in the context of RAG pipelines, including for language-based agents.